\documentclass[runningheads]{llncs}
\usepackage[T1]{fontenc}

\usepackage{graphicx}
\usepackage{amsmath,amssymb,amsfonts}
\usepackage{graphicx}
\usepackage{textcomp}
\usepackage{caption}
\usepackage{subcaption}
\usepackage{pdfpages}

\usepackage{tabularx}
\usepackage{hyperref}
\usepackage{subcaption} 
\usepackage{standalone}
\usepackage{float}
\usepackage[noend]{algpseudocode}
\usepackage{comment}
\usepackage[backend=biber, style=ieee]{biblatex}
\def\BibTeX{{\rm B\kern-.05em{\sc i\kern-.025em b}\kern-.08em
    T\kern-.1667em\lower.7ex\hbox{E}\kern-.125emX}}

\begin{document}

\title{MiCRO for Multilateral Negotiations}

\author{David Aguilera-Luzon\inst{1} \and
Dave de Jonge\inst{2} \and
Javier Larrosa\inst{1}}
\authorrunning{Aguilera-Luzon et al.}
\institute{Universitat Politècnica de Catalunya, Barcelona, Spain \\
\email{david.aguilera.luzon@upc.edu, larrosa@cs.upc.edu} \and
IIIA-CSIC, Bellaterra, Spain \\
\email{davedejonge@iiia.csic.es}}
\maketitle

\begin{abstract}
Recently, a very simple new bilateral negotiation strategy called MiCRO was introduced that does not make use of any kind of opponent modeling or machine learning techniques and that does not require fine-tuning of any parameters. Despite its simplicity, it was shown that MiCRO performs similar to---or even better than---most state-of-the-art negotiation strategies. This lead its authors to argue that the benchmark domains on which negotiation algorithms are typically tested may be too simplistic. However, one question that was left open, was how MiCRO could be generalized to multilateral negotiations. In this paper we fill this gap by introducing a multilateral variant of MiCRO. We compare it with the winners of the Automated Negotiating Agents Competitions (ANAC) of 2015, 2017 and 2018 and show that it outperforms them. Furthermore, we perform an empirical game-theoretical analysis to show that our new version of MiCRO forms an empirical Nash equilibrium.
\end{abstract}





\textbf{Index Terms}—Automated Negotiation, Multi-Agent Systems, Multilateral Negotiation, Concession Strategies, MiCRO Strategy, Negotiation Benchmarking, Best-Response Dynamics, ANAC, Strategy Robustness, Nash Equilibrium.

\section{Introduction}
\label{sec:introduction}

Negotiation is a key mechanism for conflict resolution and resource allocation in multi-agent systems, where autonomous agents must balance self-interest with cooperation. Automated negotiation focuses on designing agents that exchange proposals to reach mutually acceptable agreements, often under incomplete information and time constraints.
The core dynamic of automated negotiation involves agents proposing potential solutions or agreements, which other participating agents can then choose to accept or reject based on their own criteria and objectives~\cite{Baarslag2016,YasserMohammad,deJonge2025IntroToNego}.

Despite their fundamentally self-interested nature, a critical aspect of successful automated negotiation is the agent's ability to formulate proposals that offer sufficient value to their counterparts in order to get them accepted. Consequently, a central challenge for any negotiating agent is to strike an effective balance: maximizing its own utility while simultaneously ensuring that the proposed terms provide adequate benefits to others, thereby increasing the likelihood of acceptance and successful agreement.

This approach is commonly observed in what is known as \textit{Proposal-Based Negotiation} (PBN), a paradigm where the primary mode of communication involves agents iteratively exchanging offers and counter-offers, or deciding to terminate the negotiation process if an agreement cannot be reached \cite{Baarslag2016,deJonge2025IntroToNego,YasserMohammad}. The specific mechanics of how and when concessions are made define what is known as an agent’s \textit{strategy}.

To objectively evaluate and compare different negotiation strategies and outcomes, a variety of success metrics are employed. These include\cite{Baarslag2016,deJonge2025IntroToNego,YasserMohammad,0cf2a79a-0a79-387f-bd95-eb707896265e}:
\begin{itemize}
    \item The average utility achieved by an agent.
    \item The fairness of the outcome, often measured using the Nash bargaining solution or the Kalai-Smorodinsky solution, which offer different perspectives on equitable distribution. \cite{0cf2a79a-0a79-387f-bd95-eb707896265e,YasserMohammad}
    \item The distance of the agreed outcome to the Pareto frontier. The Pareto frontier represents outcomes where no agent can be made better off without making another worse off.\cite{6343239,6978625}
    \item The joint utility of all agents involved.
    \item The robustness of an agent’s strategy against exploitation by adversaries.
    \item The time taken until an agreement is reached (measured either in real time or by the number of exchanged messages).
\end{itemize}

Early strategies were simple heuristics like linear or conceder models. Modern strategies often include \textit{opponent modeling} inferring opponent preferences from behavior to improve outcomes, using a wide range of methodologies such as Bayessian Networks, Machine Learning, etc. \cite{deJonge2025IntroToNego,YasserMohammad,Bayesian_Hindriks,Gaussian_williams}. Opponent modeling enhances robustness and enables more efficient, mutually beneficial agreements.

Current research predominantly focuses on Proposal-Based Negotiation (PBN). Within PBN, protocols like the \textit{Alternating Offers Protocol} (AOP), where two agents take turns making offers, are standard for bilateral (two-party) negotiations. For multilateral negotiations involving more than two agents, the most widely adopted protocol is an extension called the \textit{Stacked Alternating Offers Protocol} (SAOP), where any agent may make a new proposal or accept the most recent offer made by another agent \cite{Baarslag2011_ANAC2010}.

\subsection{Benchmarking Through ANAC}
The \textbf{International Automated Negotiating Agents Competition (ANAC)} provides a standardized evaluation platform using the GENIUS environment \cite{Genius}.

The problems used as benchmarks in ANAC are called \textit{negotiation domains}\cite{Baarslag2011_ANAC2010}. A domain is defined by:
\begin{itemize}
    \item A set of \textbf{issues}, each with different discrete possible values.
    \item A set of \textbf{preference profiles}, one per agent, which determine how desirable each possible agreement is to the agent. These preferences are typically represented as utility functions that remain \textit{private and fixed} during negotiation.
\end{itemize}

The combination of issues and preferences creates a large space of possible offers. These domains can vary significantly in size, structure, and conflict level (for example how opposed the preferences of the agents are). The benchmark problems in ANAC are hand-crafted or based on real-world-inspired scenarios to test negotiation behavior under different levels of complexity\cite{Baarslag2011_ANAC2010}.

In ANAC, the agents compete in round-robin tournaments under fixed protocols and deadlines. The performance for each agent is then evaluated across multiple dimensions:
\begin{itemize}
    \item \textbf{Average utility}: Mean utility achieved by the agent over all sessions. ANAC used to grade the agents their Average Individual Utility\cite{Baarslag2011_ANAC2010}.
    \item \textbf{Agreement rate}: Fraction of sessions in which an agreement was reached.
    \item \textbf{Social welfare}: Sum of the utilities of both agents.
    \item \textbf{Fairness metrics}: Such as distance to the Nash or Kalai–Smorodinsky solution.\cite{0cf2a79a-0a79-387f-bd95-eb707896265e}
\end{itemize}

Overall, ANAC has become a critical driver in advancing the field of automated negotiation, establishing a shared experimental framework that fosters reproducibility, comparability, and continuous methodological progress.

\subsection{Scope and Objectives}
\label{sec:scope_objectives}

Within this landscape, the \textbf{MiCRO} strategy \cite{deJonge2022MiCRO} was designed to demonstrate how a remarkably simple, model-free strategy could outperform highly sophisticated agents in bilateral negotiation domains, particularly those used in past editions of ANAC. However, its effectiveness in multilateral settings remained untested.

This work extends MiCRO to multilateral domains, where increased agent count introduces new strategic complexities. We assess whether MiCRO's simplicity remains effective and whether classical ANAC benchmarks on Multilateral domains pose sufficient challenge in this broader setting.

\textbf{The main objectives of this work are:}
\begin{itemize}
    \item To adapt the MiCRO strategy to multi-agent negotiation settings and evaluate its effectiveness in domains involving more than two agents.
    \item To compare the performance of MiCRO against established agents in multi-lateral domains, including whether they lead to near-optimal or game-theoretically stable outcomes.
\end{itemize}

Ultimately, in line with de Jonge’s conclusion \cite{deJonge2022MiCRO}, this paper follows the principle that if negotiation strategies are to be evaluated using traditional ANAC-style domains then those strategies should, at minimum, be compared against MiCRO.

All source code associated with this new strategy and the experiments will be released publicly following the publication of this work.

\section{Formalization of Core Concepts in Automated Negotiation}

An automated negotiation session is a structured interaction among a set of autonomous agents $\mathcal{A} = \{a_1, a_2, \dots, a_k\}$ who exchange proposals in a shared \textbf{negotiation domain} $\Omega$ under a specific \textbf{negotiation protocol} $\mathcal{P}$. Each agent aims to maximize its own \textbf{utility function} $u_i : \Omega \rightarrow \mathbb{R}$ while constrained by $\mathcal{P}$ and limited knowledge of the other agents' preferences \cite{Genius}.

This section formalizes key concepts used in automated negotiation, especially as implemented in the ANAC competitions.

\subsection*{Negotiation Domain and Offer Space}

The \textbf{negotiation domain} $\Omega$ or \textbf{offer space} is the finite set of all possible \textbf{offers} that agents can propose. Each offer $\omega \in \Omega$ is a complete assignment of values to a set of $m$ \textbf{issues}.
Issues are the topics which the agents have to select a value, each issue $j$ has a domain $D_j$ of discrete values. Formally:

\[
\Omega = D_1 \times D_2 \times \dots \times D_m
\]

Each $\omega = (\omega_1, \omega_2, \dots, \omega_m) \in \Omega$ is an \emph{offer} that assigns one value to each issue.

\subsection*{Utility Function}

Each agent $a_i$ has a private \textbf{utility function} $u_i : \Omega \rightarrow [0,1]$ that ranks how desirable each offer is. A commonly used model is the \emph{linear additive utility function} \cite{Vente2020,BAARSLAG201373}:

\[
u_i(\omega) = \sum_{j=1}^{m} w_j^i \cdot e_j^i(\omega_j)
\]

where:
\begin{itemize}
    \item $w_j^i$ is the weight agent $i$ assigns to issue $j$, with $\sum_{j=1}^m w_j^i = 1$,
    \item $e_j^i : D_j \rightarrow [0,1]$ is the evaluation function that maps issue values to normalized utility.
\end{itemize}

\subsection*{Reservation Value}

The \textbf{reservation value} (or \textbf{reservation utility}) $r_i$ of agent $a_i$ represents the utility the agent receives when he does not make an agreement~\cite{Vente2020,deJonge2025IntroToNego}.

\subsection*{Agreement and Optimality}

An offer $\omega^* \in \Omega$ becomes an \textbf{agreement} if it is accepted by all agents under the protocol rules. A solution is \textbf{Pareto optimal} if no other offer improves one agent's utility without decreasing another's. The set of all such outcomes defines the \textbf{Pareto frontier} \cite{6343239,6978625}.

\subsection*{Agents and Strategies}

An \textbf{agent} $a_i \in \mathcal{A}$ is a computational entity that implements a negotiation \textbf{strategy} $\pi_i$, which governs its observable behavior. 

The strategy of an agent encompasses its opponent modeling (if any), concession tactics, and time or risk management. Strategy classes include~\cite{Baarslag2016}:
\begin{itemize}
    \item \emph{Holding strategy}: maintains proposals near its maximum utility.
    \item \emph{Linear concession strategy}: reduces utility aspiration linearly over time at a linear rate.
    \item \emph{Conciliatory strategy}: concedes rapidly.
\end{itemize}
More advanced strategies incorporate learning and opponent modeling to adjust $\pi_i$ dynamically.

Formally, a strategy can be represented as a function:
\[
\pi_i: \mathcal{H} \rightarrow \Omega \cup \{\text{accept}, \text{end}\}
\]
where $\mathcal{H}$ denotes the negotiation history observed up to that point~\cite{Vente2020}.

\subsection*{Negotiation Protocols $\mathcal{P}$}

A \textbf{negotiation protocol} defines the allowed actions and order of interactions. Common protocols include:

\begin{itemize}
    \item \textbf{Alternating Offers Protocol (AOP)}: Two agents take turns proposing offers.
    \item \textbf{Stacked Alternating Offers Protocol (SAOP)}: Used for \textbf{multilateral negotiation} ($|\mathcal{A}| > 2$). Agents take turns proposing offers; any agent may accept the latest offer.\cite{Aydoğan2017}
    \item \textbf{Argumentation-Based Negotiation (ABN)}: Enriches communication with justifications and rebuttals \cite{doi103233/AAC-160012, RAHWAN_RAMCHURN_JENNINGS_McBURNEY_PARSONS_SONENBERG_2003}.
\end{itemize}

\subsection*{Bilateral vs. Multilateral Negotiation}

\textbf{Bilateral negotiation} involves exactly two agents ($|\mathcal{A}|=2$). It is simpler to model and solve, but limited in strategic richness. \textbf{Multilateral negotiation} ($|\mathcal{A}| \geq 3$) is more complex due to coalition dynamics and higher risk of non-agreement. The SAOP protocol is widely used in this setting.

\subsection{ANAC Scenario Configuration (2015--2018)}

From 2015 to 2018, the ANAC competition focused on \textbf{multilateral negotiation} scenarios involving exactly \textbf{three agents} ($k = 3$). Negotiations were conducted using the \textbf{SAOP protocol}, with agents interacting over diverse domains $\Omega$ with private utility functions $u_i$.

Each ANAC scenario is defined by a triple $(\Omega, \{u_i\}, \mathcal{P})$, where:
\begin{itemize}
    \item $\Omega$ is the set of possible offers (loaded at runtime),
    \item $\{u_i\}$ are private, fixed utility functions,
    \item $\mathcal{P} = \text{SAOP}$ governs the interaction rules.
\end{itemize}

Agents were evaluated based on their \textbf{average utility} across multiple negotiation sessions. Domains varied in complexity, and scenarios were designed to test adaptability, fairness, and strategic robustness. Performance metrics included:
\begin{itemize}
    \item \textbf{Average utility} over all sessions,
    \item \textbf{Agreement rate},
    \item \textbf{Distance to Pareto frontier},
    \item \textbf{Fairness metrics}, such as Nash product \cite{Baarslag2011_ANAC2010,BAARSLAG201373}.
\end{itemize}

\section{The MiCRO Benchmark Strategy}
\label{sec:micro}
In 2022 the MiCRO (Minimal Concession in Reply to new Offers) strategy was introduced as a critique of the perceived simplicity of linear bilateral domains commonly used in the Automated Negotiating Agents Competition (ANAC) and as a new benchmark for future negotiation agent development~\cite{deJonge2022MiCRO}.  Their paper presented several experiments in
which MiCRO competed against top agents from various editions of the ANAC competition, and showed that MiCRO consistently outperformed each of
those agents. Furthermore, it was formally proven that under certain conditions that typically hold in the ANAC competitions, MiCRO is theoretically optimal among all so-called `consistent' strategies \cite{deJonge2024theoreticalPropertiesMicro}.

MiCRO also participated itself in the ANAC 2022 competition. While it only
ended in 9th place out of 19 participants, it was later shown \cite{renting} that it was in fact the
best participant from a game-theoretical point of view. That is, it was shown to form
the best empirical Nash equilibrium among all submitted agents. Furthermore, MiCRO was submitted again to ANAC in 2023 and ended in second place,
out of 15 participants. These results are remarkable, given that, unlike virtually any other strong ANAC participant, MiCRO does not apply any form of opponent modeling or machine learning techniques. Furthermore, MiCRO does not depend on any parameters that need to be fine-tuned.

From these results, the authors of \cite{deJonge2022MiCRO}
concluded that the traditional ANAC domains, despite being widely used, might be too simplistic to truly differentiate the capabilities of complex negotiation algorithms, particularly those employing sophisticated opponent modeling techniques. Note, however, that they insist that MiCRO is not intended as a strategy for practical applications but rather serves as an analytical tool to assess the complexity of domains and the relative strength of other negotiation strategies.

\subsection{How MiCRO Works}\label{sec:how_micro_works}

Simplifying, MiCRO works as follows: 
before the negotiations start, it creates a list that contains all offers from the offer space, sorted in descending order of utility for the agent itself.

Whenever it's the  MiCRO-agent's turn, and it has so far not made more unique proposals than its opponent, then it proposes a new offer. Specifically, it proposes the next offer on its list, so it proposes the one with the highest utility that it has not yet proposed before. If, on the other hand, the MiCRO-agent has already made more proposals than the opponent, it simply repeats a previous offer selected at random from those it has already proposed.

This behavior ensures that MiCRO always makes the smallest possible concession in response to a new offer. Importantly, the strategy does not require any knowledge of the opponent’s preferences and therefore does not employ any form of opponent modeling. Moreover, it does not even need to know its own utility function exactly. It only needs a complete preference ordering over the domain's offers.

\subsubsection*{Concession Strategy}
The core mechanism of MiCRO is its concession strategy:
\begin{enumerate}
    \item \textbf{Offer Sorting:} Before negotiation begins, MiCRO creates a list of all possible offers in the domain ($\Omega$), sorted in descending order of utility for itself. Let this sorted list be $(\omega_{1}, \omega_{2}, ..., \omega_{K})$, where $u_{1}(\omega_{1}) \ge u_{1}(\omega_{2}) \ge \dots \ge u_{1}(\omega_{K})$ and $K = |\Omega|$. Notably, MiCRO does not need to know its actual utility function values; it only requires a complete preference ordering over the offers.
    \item \textbf{Proposal Mechanism:} When it is MiCRO's turn to make a proposal, its action depends on the history of offers received and made:
    \begin{itemize}
        \item Let $n$ be the count of \textit{different} offers received from the opponent so far.
        \item Let $m$ be the count of \textit{different} offers MiCRO has proposed so far.
        \item If $m \le n$ (meaning the opponent has proposed at least as many new offers as MiCRO), MiCRO proposes the next offer on its sorted list that it has not yet proposed, i.e., $\omega_{m+1}$. This is considered its ``minimal concession.''
        \item If $m > n$ (meaning MiCRO has proposed more unique offers than it has received), MiCRO repeats one of its previously proposed offers by picking a random integer $r$ such that $1 \le r \le m$ and proposing $\omega_{r}$.
    \end{itemize}
    \item \textbf{Reservation Value Handling:} If proposing $\omega_{m+1}$ would result in an offer below MiCRO's reservation value ($u_{1}(\omega_{m+1}) < rv_{1}$), it will instead repeat a random previous proposal, even if $m \le n$.
\end{enumerate}

\subsubsection*{Acceptance Strategy}
MiCRO applies the following acceptance strategy:
Let $\omega_{low}$ be the lowest utility offer MiCRO is willing to propose at that time (defined as $\omega_{m+1}$ if $m \le n$, and $\omega_{m}$ if $m > n$). MiCRO will accept a received offer $\omega$ if and only if $u(\omega) \ge \max\{u(\omega_{low}), rv\}$, where $u$ is its utility function and $rv$ its reservation value.

\section{New MiCRO Strategy for Multi-lateral Scenarios}


The goal of this work is to extend the existing MiCRO strategy to enable it to also negotiate in \textit{multi}-lateral negotiation scenarios. This involves leveraging the existing negotiation protocol while modifying the decision-making process to consider all offers made by all participating agents, rather than just one opponent. The aim is to retain the original behavior of MiCRO while making it applicable to the broader multi-agent context.


To achieve this, the key question becomes: \textit{how should MiCRO determine when to propose a new offer based on the multiple opponents it faces?}

Recall from Section \ref{sec:how_micro_works} that MiCRO proposes a new offer when $m\leq n$, where $m$ is the number of unique offers it already proposed itself, and $n$ the number of unique offers proposed by the opponent. So, to generalize this to \textit{multilateral} negotiations, the question is how to replace the definition of $n$. For this, we initially considered the following three options:
\begin{itemize}
    \item \textbf{MiCRO-Max:} Proposes a new offer whenever $m \leq n_{max}$, where $n_{max}$ is defined as:\begin{equation}\label{eq:n_max}
    n_{max} := \max\{n_1, n_2, \dots, n_k\}.
    \end{equation}
    and where $n_i$ denotes the number of offers made by opponent $i$, and there are $k$ opponents.
    \item \textbf{MiCRO-Min:}  Proposes a new offer whenever $m \leq n_{min}$, where $n_{min}$ is defined as:
    \begin{equation}\label{eq:n_min}
        n_{min} := \min\{n_1, n_2, \dots, n_k\}.
    \end{equation}
    \item \textbf{MiCRO-Mean:} Proposes a new offer whenever $m \leq n_{mean}$, where $n_{mean}$ is defined as:\begin{equation}\label{eq:n_mean}
    n_{mean} := \frac{1}{k} \sum_{i=1}^{k} n_i.
    \end{equation}
\end{itemize}

Note that each of these variants indeed generalizes MiCRO in the sense if $k=1$, they all become identical to the original MiCRO.

At first sight, each of these three variants may seem a reasonable generalization of MiCRO. However, in this paper we argue that MiCRO-Min is actually the only reasonable one.

The problem with the other two algorithms, is that they are easily exploited. This is most clearly illustrated in the case that there are three agents in total, of which two employ the MiCRO-Max strategy, and the third agent is a very hardheaded agent that is not willing to concede much. In this case, the two instances of MiCRO-Max will keep making concessions, ignoring the fact that the third agent does not make any new proposals. This means they could end up proposing some offer that yields very  high utility for the third agent, but very low utility for the two MiCRO-Max agents themselves. 
This is in direct conflict with the philosphy of the original MiCRO strategy to never concede more than the opponent. 

A similar situation could occur with the MiCRO-Mean strategy. On the other hand, the MiCRO-Min strategy does not suffer from this problem, as it requires \textit{all} agents to make at least as many new proposals as itself. For this reason we will focus in this paper only on the MiCRO-Min strategy and consider it as the only correct generalization of MiCRO.

Note that even if the other two versions may in practice still  perform well, this does not change the fact that they contradict the basic idea of the original MiCRO to never concede more than the opponent, so we still could not consider them to be `correct' generalizations of MiCRO.



\subsection{The Deadlock Problem.}
\label{subsec:micro-deadlock}

Above, we have argued that MiCRO-Max and MiCRO-Mean do not correctly generalize MiCRO. Here, we furthermore argue that there is another problem with the multilateral versions of MiCRO proposed above. We call this the `deadlock problem' and again it is easiest explained with an example.

Suppose there are three agents, denoted $a_1$, $a_2$ and $a_3$ that each apply the MiCRO-Min strategy. Furthemore, assume that it is currently the turn of $a_1$ and that we have $n_2 < n_3 = n_1$ where each $n_i$ is the number of unique offers so far proposed by agent $a_i$. The negotiations proceed as follows:
\begin{enumerate}
    \item It is $a_1$'s turn. Since $n_1 > \min \{n_2,n_3\}$,
agent $a_1$ will not propose any new offer. Instead  $a_1$ will repeat some offer $\omega$ that he already proposed before. This means the number $n_1$ will not increase.
    \item Next, it is  $a_2$'s turn. Since $n_2 \leq \min \{n_1, n_3\}$, agent $a_2$ is now willing to propose a new offer $\omega'$. However, let us assume that she actually prefers the offer $\omega$ that was just proposed by $a_1$.
    That is, we have $u_2(\omega') < u_2(\omega)$. This means that instead of proposing $\omega'$ she will instead \textit{accept} offer $\omega$. This implies that $n_2$ also does not increase. 
    \item Next, it is $a_3$'s turn. Let us assume that $a_3$ rejects the offer $\omega$.
    Furthermore, since $n_3 > \min \{n_1,n_2\}$,
agent $a_3$ will also not propose any new offer. Instead, just like $a_1$, she will repeat some earlier offer and $n_3$ will not increase.
\end{enumerate}
We see that in this way 
none of the three agents makes any new proposals, and thus none of the three numbers $n_1$, $n_2$ and $n_3$ changes.
In this way, it is possible that the agents get stuck in a deadlock, in which $a_1$ and $a_3$ keep repeating old proposals and $a_2$ keeps accepting the proposals by $a_1$, which, however, keep getting rejected by $a_3$




To solve this issue, we just needed to make a small change to the definition of MiCRO-Min. Specifically, we had to redefine the numbers $n_{i}$ to also include the number of acceptances. Specifically, for any agent $a_i$ we define:
\begin{equation}\label{eq:n_i}
    n_i := \{\omega \in \Omega \mid \omega \text{\ has\ been\ either\ proposed\ or\ accepted\ by\ agent\ } a_i\}
\end{equation}
With this new definition  the deadlock does not occur, because now, in the example above, whenever agent $a_2$ accepts the offer proposed by $a_1$, this increases the value of $n_2$, which means that we will next have $n_1 = n_2 = n_3$, and thus that $a_1$ and $a_3$ will again be willing to propose new offers.\footnote{Note that with this new definition, if all agents play MiCRO-Min then we always have $|n_i-n_j| \leq 1$, so if $n_2 < n_3$ then we must have $n_2 = n_3-1$, and therefore, after $a_2$ accepts we will indeed have $n_2 = n_3$}

The idea is that in the SAOP, proposing an offer is basically the same as accepting an offer. So, whenever an agent \textit{accepts} a new offer it can just as much be seen as a concession as when it \textit{proposes} a new offer. Therefore, acceptances should be counted equally as proposals.

Furthermore, note that for the original, bilateral, version of MiCRO it was not necessary to count acceptances, because in the bilateral AOP, the negotiations are finished anyway as soon as one of the agents accepts a proposal.

In the rest of this paper, we will simply refer to our new agent, based on Eq.~(\ref{eq:n_min}) and Eq.~(\ref{eq:n_i}), as `MiCRO', since it will be clear from context whether we are talking about the original bilateral variant, or our new multilateral variant, and because the bilateral variant can be seen as a special case of the multilateral variant anyway.

\section{Experimental Setup}
\label{sec:experimental_setup}

To evaluate the performance of our new multilateral version of MiCRO, we have performed a series of experiments. All these experiments were conducted on the NegMAS framework, in combination with the so-called `\textit{Genius bridge}' which allows agents from the older Genius framework to be run from inside  NegMas.

All experiments were conducted on a MacBook equipped with an Apple M4 Pro chip, 24GB of RAM, running macOS Sequoia 15.1.


\subsection{Selection of Opponent Agents}
\label{subsec:opponent_selection}
We tested our multilateral version of MiCRO against the winners of ANAC 2015, and 2017, and 2018, because these were the years in which the competition involved multilateral negotiations. ANAC 2016 also involved multilateral negotiations, but the winner of that year (`Caduceus') was not available to be run in NegMas through the Genius bridge, so we could not include it in our experiments.


Specifically, we used the following three agents:
\begin{itemize}
    \item \textbf{Atlas3}: Winner of ANAC 2015 \cite{Mori2017Atlas3}.
    \item \textbf{PonPoko}: Winner of ANAC 2017
    \item \textbf{AgreeableAgent2018}: Winner of ANAC 2018 \cite{Mirzayi2022AgreeableAgent}.    
\end{itemize}


Subsequent to ANAC 2018, the competition's focus shifted back towards bilateral negotiation scenarios. Therefore, we could not use any of the agents from the later competitions.


\subsection{Negotiation Scenarios}
\label{subsec:scenario_selection}

For the negotiation scenarios (domains and preference profiles), we adopted the following suite of scenarios utilized in the ANAC 2015 competition:\\
\texttt{group1-university}, \texttt{group2-dinner}, \texttt{group2-politics},
\texttt{group3-bank\_robbery}, \texttt{group5-car\_domain}, \texttt{group6-tram},
\texttt{group8-holiday}, \texttt{group9-killer\_robot}, \texttt{group9-vacation}, \texttt{group11-car\_purchase}.

They were directly available to us in the Genius framework. At the time the tests were performed the other scenarios from ANAC 2015 were not available to use due to some error in their structure.


\subsection{Test Methodologies}
\label{subsec:test_methodologies}
The battery of tests involved running iterations across all negotiation scenarios mentioned above. Just like in the multilateral ANAC competitions, each negotiation involved three agents. For each scenario, every possible combination of agents was tested. It was specified that agents could be repeated within a session (for instance a session could consist of three MiCRO agents) and that the order of agents within a combination did not matter for defining a unique test case.
For $n=4$ distinct agent types and $k=3$ agents per session, the number of unique combinations is:
$$ \binom{4+3-1}{3} = \binom{6}{3} = \frac{6 \times 5 \times 4}{3 \times 2 \times 1} = 20 $$

Some of the scenarios contained more than three utility functions, so for those cases we simply selected the first three utility functions to be used in our experiments. In order to ensure fairness, we tested all the agents with all the three utility functions.
So for the tests each one of these utility function  was tested on each agent of the triplet.
In this case, no repetition was allowed (the agents need a unique utility function) so for each triplet distributing 3 utility functions, means that there are 6 possible combinations. This resulted that for each scenario, all the 20 triplets negotiated 6 times between each other.


\subsection{Performance Metrics}
In order to assess the performance of MiCRO we have measured the following metrics.
\begin{itemize}
    \item \textbf{Mean Utility}: This is the primary performance indicator and represents the agent’s average utility across all negotiation sessions. It is the main value used to compare agents' effectiveness.

    \item \textbf{Standard Error}: This metric reflects how accurately we have measured the mean utility. The smaller it is, the more reliable the mean utility is.
    
    \item \textbf{Utility on Agreement}: This metric reports the agent’s average utility only in those sessions where an agreement was reached. It helps explain whether an agent concedes more or less when it does reach agreement, but it should not be used by itself to assess performance since it ignores failed negotiations.
    
    \item \textbf{Agreement Rate}: This value indicates the percentage of negotiation sessions that ended in an agreement. Like `utility on agreement', it is not a direct measure of quality, but it helps to contextualize the mean utility.

\end{itemize}

\section{Results}

The results of our experiments are displayed in Table~\ref{tab:utility_statistics}. Note that Mean Utility, Standard Error and Utility on Agreement, have all been multiplied by a factor of 100 for the sake of readability.

\begin{table}[h]
    \centering 
    \begin{tabular}{l|c|c|c}
        \textbf{Agent} & \textbf{Mean Utility \textpm  Std. Err.} & \textbf{Utility on Agreement} & \textbf{Agreement Rate} \\
        \hline
        MiCRO & 81.40 \textpm 0.35 & 87.53 & 92.62\% \\
        Agreeable & 80.46 \textpm 0.32 & 81.70 & 98.33\% \\
        PonPoko & 79.59 \textpm 0.40 & 88.07 & 89.64\% \\
        Atlas3 & 75.84 \textpm 0.31 & 76.51 & 99.00\% \\
    \end{tabular}
    \caption{Utility statistics for each agent, sorted by Mean Utility}\label{tab:utility_statistics}
\end{table}



We see that that MiCRO outperforms the three other agents in terms of mean utility, although it should be noted that the difference between MiCRO and AgreeableAgent is relatively small compared to their standard errors, so this difference may not be statistically significant. Nevertheless, these results are very important, because unlike the other three agents, MiCRO does not use any form of opponent modeling or machine learning techniques.


From the other two metrics we see that MiCRO is a tougher agent than AgreeableAgent and Atlas3, since it makes less agreements, but it scores higher utility when it does make an agreement. On the other hand, it is not as tough as PonPoko, which scores higher utility on agreement than MiCRO, but makes less agreements than MiCRO. The fact that MiCRO scores the highest mean utility means that, overall, MiCRO strikes the best balance between Utility on Agreement and Agreement Rate.


\subsection{Game-Theoretic Analysis of MiCRO as a Strategic Choice}
\label{sec:game_theory_micro}
To gain deeper insight into the strategic robustness and adaptability we conducted a game-theoretic analysis grounded in best-response dynamics.
In our analysis, we imagine a setting in which three players must simultaneously select a strategy from the set \{Agreeable, PonPoko, Atlas3, MiCRO\}. The expected utility of each strategy combination is given by the \textbf{mean utility} score obtained in our experiments.

For each triplet of players, we fix two strategies and allow the third player to choose the optimal strategy in response to the two other players. Table~\ref{tab:agents_best_pairs} displays the expected utility that player 3 can then achieve. For example, we see that if the first two players each choose AgreeableAgent, and the third player chooses MiCRO, then the third player will achieve an expected utility of 0.8605. On the other hand, if the third player chooses PonPoko, then he will achieve an expected utility of 0.9181. Since we see that this is the highest utility he can expect to achieve when the other two players choose AgreeableAgent, we say that PonPoko is the best response against (AgreeableAgent, AgreeableAgent).


\begin{table}[h]
\centering
\resizebox{0.5\textwidth}{!}{
\begin{tabular}{|c|c|c|}
\hline
\textbf{Opponent Pair} & \textbf{Agent} & \textbf{Average Utility} \\
\hline
(Agreeable, Agreeable) & PonPoko & 0.9181 \\
                       & Micro & 0.8605 \\
                       & Agreeable & 0.8152 \\
                       & Atlas3 & 0.7224 \\
\hline
(Agreeable, Atlas3) & PonPoko & 0.9183 \\
                    & Micro & 0.8912 \\
                    & Agreeable & 0.8819 \\
                    & Atlas3 & 0.6998 \\
\hline
(Agreeable, Micro) & Micro & 0.8068 \\
                   & PonPoko & 0.8059 \\
                   & Agreeable & 0.7587 \\
                   & Atlas3 & 0.7011 \\
\hline
(Agreeable, PonPoko) & Micro & 0.8387 \\
                     & PonPoko & 0.8195 \\
                     & Agreeable & 0.7611 \\
                     & Atlas3 & 0.7203 \\
\hline
(Atlas3, Atlas3) & Agreeable & 0.9945 \\
                 & PonPoko & 0.9526 \\
                 & Micro & 0.9128 \\
                 & Atlas3 & 0.8513 \\
\hline
(Atlas3, Micro) & Agreeable & 0.8693 \\
                & Micro & 0.8692 \\
                & PonPoko & 0.8058 \\
                & Atlas3 & 0.8021 \\
\hline
(Atlas3, PonPoko) & Micro & 0.8380 \\
                  & PonPoko & 0.8302 \\
                  & Agreeable & 0.8118 \\
                  & Atlas3 & 0.7591 \\
\hline
(Micro, Micro) & Micro & 0.8369 \\
               & Atlas3 & 0.7591 \\
               & Agreeable & 0.7165 \\
               & PonPoko & 0.6146 \\
\hline
(Micro, PonPoko) & Agreeable & 0.7005 \\
                 & PonPoko & 0.6927 \\
                 & Atlas3 & 0.6876 \\
                 & Micro & 0.6313 \\
\hline
(PonPoko, PonPoko) & PonPoko & 0.7460 \\
                   & Micro & 0.7435 \\
                   & Agreeable & 0.7267 \\
                   & Atlas3 & 0.7091 \\
\hline
\end{tabular}
}
\caption{Average Utility of each agent against every possible pair of opponents}
\label{tab:agents_best_pairs}
\end{table}

\paragraph{Best Response Frequency:}
The number of times each agent appears as the best response:  \textbf{MiCRO}: 4 times, \textbf{PonPoko}: 3 times, \textbf{Agreeable}: 3 times, \textbf{Atlas3}: 0 times.

\begin{figure}[h]
    \centering
    \includegraphics[width=1\textwidth]{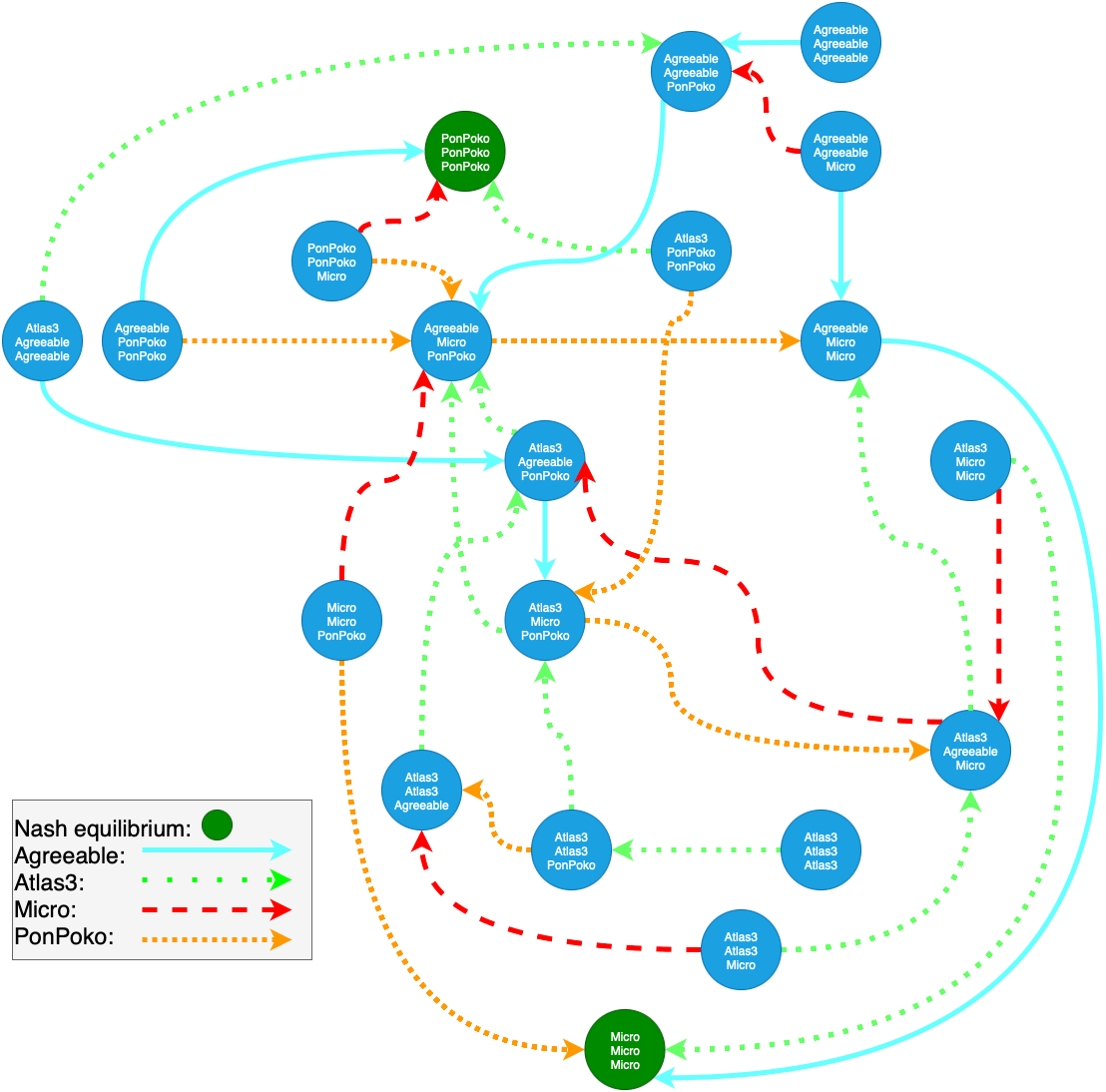}
    \caption{Best-response dynamics across all triplets of strategy profiles. Arrows indicate unilateral improvements; Nash equilibria are highlighted.}
    \label{fig:best_response_dynamics}
\end{figure}

Based on these data, we have generated a best response graph, which is displayed in Figure~\ref{fig:best_response_dynamics}.
In this graph:
\begin{itemize}
    \item \textbf{Nodes} represent a specific combination of strategies selected by the three players, for example: \texttt{(Player 1: Atlas3, Player 2: PonPoko, Player 3: MiCRO)}.
    \item \textbf{Directed Edges} (colored arrows) illustrate a player's \textit{best response}. An arrow from one node to another signifies that the corresponding player can increase their utility by unilaterally changing their strategy, thus moving the game to the new node. The color and shape of the arrows represent the strategy that is being changed. For example, we see that there is a green arrow going from the node $\{Atlas3, Agreeable, Agreeable\}$ to the node  $\{PonPoko, Agreeable, Agreeable\}$. This means that if one player chooses Atlas3, while its two opponents each choose AgreeableAgent, then that agent can increase its expected utility by switching to PonPoko.
    \item If a node does not have any outgoing edges, it means that no player can increase his utility by changing strategy. In other words, it is a \textbf{Nash Equilibrium}. We have highlighted these nodes in green.
\end{itemize}

This pairwise best-response analysis reinforces the overall strategic strength of \textbf{MiCRO}, which appears the most as the best choice. Its adaptability and robust performance across varying opponents reflect its suitability as a default negotiation agent in multilateral environments.

\textbf{PonPoko} also performs competitively, particularly against more cooperative or less assertive configurations (such as Agreeable-based pairs), while \textbf{Agreeable} dominates against weaker opponents like Atlas3. Notably, \textbf{Atlas3} never emerges as a best response, indicating its relatively lower strategic viability under competitive conditions.

It is important to note that Nash equilibrium is obtained only in two scenarios: \{MiCRO, MiCRO, MiCRO\} and \{PonPoko, PonPoko, PonPoko\}. However, a comparison of their respective self-play utilities provides additional insight: three \textbf{PonPokos} achieve an average utility of 0.7460 each, whereas three \textbf{MiCROs} achieve a significantly higher average utility of 0.8369 each. This difference indicates that if all agents were to choose MiCRO, they would be collectively better off, implying that \{MiCRO, MiCRO, MiCRO\} is a more favorable Nash equilibrium than \{PonPoko, PonPoko, PonPoko\}.




Overall, the structure of best-response dynamics supports the conclusion that \textbf{MiCRO} is a strategically dominant or at least highly competitive choice in multilateral settings. The graphs reinforce this conclusion by showing that \textbf{MiCRO} performs well against most state-of-the-art strategies, frequently appearing in equilibrium configurations and resisting displacement by best-response adaptations. From a game-theoretic perspective, \textbf{MiCRO} consistently constitutes part of an equilibrium in multilateral scenarios, providing a formal justification for recommending it as a benchmark strategy when evaluating new negotiation agents in multi-agent environments.

\section{Conclusions}

This work investigated the applicability of the MiCRO strategy in multilateral negotiation environments. Building on prior bilateral results \cite{deJonge2022MiCRO}, we introduced a multilateral version of MiCRO and evaluated its performance against top agents from past ANAC competitions.

The results demonstrate that:
\begin{itemize}
    \item \textbf{MiCRO is highly competitive}, matching or outperforming complex adaptive agents in multilateral settings.
    \item \textbf{MiCRO forms an empirical Nash equilibrium} among the winners of ANAC 2015, 2017 and 2018, suggesting both practical reliability and theoretical soundness under best-response dynamics.
    \item 
    \textbf{The multilateral ANAC domains can still be tackled without any form of opponent modeling}.
    \item Benchmarking systems must include richer, more adversarial, and less predictable domains to better differentiate strategy sophistication.
\end{itemize}

We have shown that the our new version of MiCRO can achieve high performance in complex multi-agent environments. 

The experiments  validate MiCRO’s adaptability and emphasize a broader concern: the continuing limitations of existing benchmarks like ANAC 2015 in evaluating negotiation strategies. The results underscore the need for evaluation protocols that can distinguish between genuine strategic sophistication and reactive efficiency.

In addition, the success of such a minimal strategy raises broader concerns about the benchmarking standards used in ANAC competitions. This calls into question the strategic depth and adversarial richness of these environments, and their effectiveness in evaluating genuinely sophisticated negotiation strategies.

Therefore, this work not only highlights MiCRO’s surprising adaptability and competitiveness, but also underscores the need for more challenging, dynamic, and strategically diverse benchmarks in automated negotiation research.

\subsection{Future Work}

Several promising directions emerge from this study:

\begin{itemize}
    \item \textbf{Domain complexity analysis:} A systematic study of the structural properties of ANAC negotiation domains would help identify which domains are genuinely challenging and which ones are susceptible to heuristic exploitation.
    \item \textbf{Robustness testing against adversarial agents:} While MiCRO has shown resistance to manipulation, further stress-testing it in environments with colluding agents or adversarial pacing strategies would help quantify its resilience under worst-case conditions.
    \item \textbf{Improved benchmarking platforms:} The limitations observed in ANAC suggest a need for new benchmarking standards that include more dynamic and asymmetric domains, strategic uncertainty, and mechanisms for testing robustness. These could be incorporated into future versions of GENIUS or NegMAS.
\end{itemize}

In conclusion, this paper highlights not only the enduring value of simple
strategies like MiCRO in automated negotiation but also the ongoing need to
critically reassess our evaluation frameworks. The goal is not merely to build
better agents, but to ensure that the environments used to test them genuinely
reflect the complexity and unpredictability of real-world negotiation.
\section*{Acknowledgments}
This work has been partially financed by INTEL Corporation, "Intel UFunding ID 14780". Also, this work has been partially financed by Grant  PID2024-157044OB-C32, funded by MICIU /AEI
/10.13039/501100011033 / FEDER, UE and partially financed by Grant PID2021-122830OB-C43, funded by MCIN/AEI/10.13039/501100011033, by “ERDF: A way of making Europe”, and by a JAE-Intro-ICU grant, JAEICU\_24\_04252, from the Spanish Scientific Research Council (CSIC).
\printbibliography 

@article{Vente2020,
  author    = {Vente, Sam and Kimmig, Angelika and Preece, Alun and Cerutti, Federico},
  title     = {The current state of automated negotiation theory: a literature review},
  journal   = {arXiv preprint arXiv:2004.02614v2 [cs.AI]},
  year      = {2020},
  url       = {https://arxiv.org/abs/2004.02614v2}
}

@inproceedings{deJonge2022MiCRO,
  author    = {Dave de Jonge},
  title     = {An Analysis of the Linear Bilateral ANAC Domains Using the {MiCRO} Benchmark Strategy},
  booktitle = {Proceedings of the AAAI Conference on Artificial Intelligence}
}

@Article{deJonge2024theoreticalPropertiesMicro,
author ={de Jonge, Dave},
title = {Theoretical Properties of the MiCRO Negotiation Strategy},
journal = {Autonomous Agents and Multi-Agent Systems},
publisher={Springer},
volume = {38},
number = {46},
year = {2024},
issn = {1573-7454},
doi = {10.1007/s10458-024-09678-1},
url = {https://doi.org/10.1007/s10458-024-09678-1}
}

@unpublished{renting,
author = {Renting, Bram and de Jonge, Dave and Hoos, Holger and Jonker, Catholijn},
title  = {Analysis of Learning Agents in Automated Negotiation},
note = {Under review}
}

@Article{Baarslag2016,
author={Baarslag, Tim
and Hendrikx, Mark J. C.
and Hindriks, Koen V.
and Jonker, Catholijn M.},
title={Learning about the opponent in automated bilateral negotiation: a comprehensive survey of opponent modeling techniques},
journal={Autonomous Agents and Multi-Agent Systems},
year={2016},
month={Sep},
day={01},
volume={30},
number={5},
pages={849-898},
issn={1573-7454},
doi={10.1007/s10458-015-9309-1},
url={https://doi.org/10.1007/s10458-015-9309-1}
}

@inproceedings{Bayesian_Hindriks,
author = {Hindriks, Koen and Tykhonov, Dmytro},
year = {2008},
month = {05},
pages = {331-338},
title = {Opponent modelling in automated multi-issue negotiation using Bayesian learning},
volume = {1},
journal = {Proceedings of the 7th International Joint Conference on Autonomous Agents and Multiagent Systems}
}

@article{Gaussian_williams,
author = {Williams, Colin and Robu, Valentin and Gerding, Enrico and Jennings, Nicholas},
year = {2011},
month = {01},
pages = {},
title = {Using Gaussian Processes to Optimise Concession in Complex Negotiations against Unknown Opponents},
journal = {IJCAI International Joint Conference on Artificial Intelligence},
doi = {10.5591/978-1-57735-516-8/IJCAI11-080}
}

@inbook{Baarslag2011_ANAC2010,
author = {Baarslag, Tim and Hindriks, Koen and Jonker, Catholijn and Kraus, Sarit and Lin, Raz},
year = {2012},
month = {01},
pages = {113-135},
title = {The First Automated Negotiating Agents Competition (ANAC 2010)},
volume = {383},
isbn = {978-3-642-24696-8},
journal = {Studies in Computational Intelligence},
doi = {10.1007/978-3-642-24696-8_7}
}

@ARTICLE{Genius,
  author = {Raz Lin and Sarit Kraus and Tim Baarslag and Dmytro Tykhonov and Koen Hindriks and Catholijn M. Jonker},
  title = {Genius: An Integrated Environment for Supporting the Design of Generic Automated Negotiators},
  journal = {Computational Intelligence},
  year = {2014},
  volume = {30},
  pages = {48--70},
  number = {1},
  doi = {10.1111/j.1467-8640.2012.00463.x},
  issn = {1467-8640},
  publisher = {Blackwell Publishing Inc},
  url = {http://dx.doi.org/10.1111/j.1467-8640.2012.00463.x}
}

@Inbook{Aydoğan2017,
author="Aydo{\u{g}}an, Reyhan
and Festen, David
and Hindriks, Koen V.
and Jonker, Catholijn M.",
editor="Fujita, Katsuhide
and Bai, Quan
and Ito, Takayuki
and Zhang, Minjie
and Ren, Fenghui
and Aydo{\u{g}}an, Reyhan
and Hadfi, Rafik",
title="Alternating Offers Protocols for Multilateral Negotiation",
bookTitle="Modern Approaches to Agent-based Complex Automated Negotiation",
year="2017",
publisher="Springer International Publishing",
address="Cham",
pages="153--167",
isbn="978-3-319-51563-2",
doi="10.1007/978-3-319-51563-2_10",
url="https://doi.org/10.1007/978-3-319-51563-2_10"
}

@book{deJonge2025IntroToNego,
title = "Introduction to Automated Negotiation",
author = "de Jonge, Dave",
year = "2025",
publisher = "self-published",
address = "Barcelona, Spain",
url = "https://www.iiia.csic.es/~davedejonge/intro_to_nego"
}

@misc{YasserMohammad, title="Yasser Mohammad Youtube Channel", url="https://www.youtube.com/@YasserMohammadElmorsy", journal={YouTube}, author="Yasser Mohammad", publisher = "self-published", year="2020"}

@article{RAHWAN_RAMCHURN_JENNINGS_McBURNEY_PARSONS_SONENBERG_2003, title={Argumentation-based negotiation}, volume={18}, DOI={10.1017/S0269888904000098}, number={4}, journal={The Knowledge Engineering Review}, author={RAHWAN, IYAD and RAMCHURN, SARVAPALI D. and JENNINGS, NICHOLAS R. and McBURNEY, PETER and PARSONS, SIMON and SONENBERG, LIZ}, year={2003}, pages={343–375}}

@article{BAARSLAG201373,
title = {Evaluating practical negotiating agents: Results and analysis of the 2011 international competition},
journal = {Artificial Intelligence},
volume = {198},
pages = {73-103},
year = {2013},
issn = {0004-3702},
doi = {https://doi.org/10.1016/j.artint.2012.09.004},
url = {https://www.sciencedirect.com/science/article/pii/S0004370212001105},
author = {Tim Baarslag and Katsuhide Fujita and Enrico H. Gerding and Koen Hindriks and Takayuki Ito and Nicholas R. Jennings and Catholijn Jonker and Sarit Kraus and Raz Lin and Valentin Robu and Colin R. Williams}
}

@article{doi103233/AAC-160012,
author = {Ana Casali and Pablo Pilotti and Carlos Chesñevar},
title ={Assessing communication strategies in argumentation-based negotiation agents equipped with belief revision1},

journal = {Argument \& Computation},
volume = {7},
number = {2-3},
pages = {175-200},
year = {2016},
doi = {10.3233/AAC-160012},

URL = {   
        https://doi.org/10.3233/AAC-160012
},
eprint = {     
        https://doi.org/10.3233/AAC-160012    
}
}

@ARTICLE{6343239,
  author={Hu, Xiao-Bing and Wang, Ming and Di Paolo, Ezequiel},
  journal={IEEE Transactions on Cybernetics}, 
  title={Calculating Complete and Exact Pareto Front for Multiobjective Optimization: A New Deterministic Approach for Discrete Problems}, 
  year={2013},
  volume={43},
  number={3},
  pages={1088-1101},
  doi={10.1109/TSMCB.2012.2223756}}

@INPROCEEDINGS{6978625,
  author={Kakimoto, Shinji and Fujita, Katsuhide},
  booktitle={2014 IEEE 7th International Conference on Service-Oriented Computing and Applications}, 
  title={Estimating Pareto Fronts Using Issue Dependency for Bilateral Multi-issue Closed Nonlinear Negotiations}, 
  year={2014},
  volume={},
  number={},
  pages={289-293},
  doi={10.1109/SOCA.2014.30}}

@article{0cf2a79a-0a79-387f-bd95-eb707896265e,
 ISSN = {00129682, 14680262},
 URL = {http://www.jstor.org/stable/1914280},
 author = {Ehud Kalai and Meir Smorodinsky},
 journal = {Econometrica},
 number = {3},
 pages = {513--518},
 publisher = {[Wiley, Econometric Society]},
 title = {Other Solutions to Nash's Bargaining Problem},
 urldate = {2025-06-21},
 volume = {43},
 year = {1975}
}

@Inbook{Mori2017Atlas3,
author="Mori, Akiyuki
and Ito, Takayuki",
editor="Fujita, Katsuhide
and Bai, Quan
and Ito, Takayuki
and Zhang, Minjie
and Ren, Fenghui
and Aydo{\u{g}}an, Reyhan
and Hadfi, Rafik",
title="Atlas3: A Negotiating Agent Based on Expecting Lower Limit of Concession Function",
bookTitle="Modern Approaches to Agent-based Complex Automated Negotiation",
year="2017",
publisher="Springer International Publishing",
address="Cham",
pages="169--173",
isbn="978-3-319-51563-2",
doi="10.1007/978-3-319-51563-2_11",
url="https://doi.org/10.1007/978-3-319-51563-2_11"
}

@article{Mirzayi2022AgreeableAgent,
  author       = {Sahar Mirzayi and
                  Fattaneh Taghiyareh and
                  Faria Nassiri Mofakham},
  title        = {An opponent-adaptive strategy to increase utility and fairness in
                  agents' negotiation},
  journal      = {Appl. Intell.},
  volume       = {52},
  number       = {4},
  pages        = {3587--3603},
  year         = {2022},
  url          = {https://doi.org/10.1007/s10489-021-02638-2},
  doi          = {10.1007/S10489-021-02638-2},
  bibsource    = {dblp computer science bibliography, https://dblp.org}
}
\end{document}